\documentclass[journal,comsoc]{IEEEtran}

\usepackage[T1]{fontenc}
\usepackage{xcolor}
\usepackage{fancyhdr}
\usepackage{tgschola}
\usepackage{lastpage}
\usepackage{amsthm,amssymb,amsmath,mathtools,fixmath}
\usepackage{amsbsy}
\usepackage{moreverb}
\usepackage{amsfonts}
\usepackage{graphicx}
\graphicspath{{SimulationFigures/}}
\usepackage{tikz}
\usepackage{listings}
\usepackage{color}
\usepackage{latexsym}
\usepackage{subfigure}
\usepackage{animate}
\usepackage{nopageno}
\colorlet{linkequation}{blue}
\usepackage[linesnumbered,ruled]{algorithm2e}
\usepackage{soul}

\def\BibTeX{{\rm B\kern-.05em{\sc i\kern-.025em b}\kern-.08em T\kern-.1667em\lower.7ex\hbox{E}\kern-.125emX}}

\usepackage{geometry}
\usepackage{lipsum}

\DeclareMathOperator*{\maxi}{maximize}
\makeatletter
\def\BState{\State\hskip-\ALG@thistlm}
\makeatother

\usepackage{cite}

\ifCLASSINFOpdf
\else
\fi
\usepackage{amsmath}
\usepackage[cmintegrals]{newtxmath}
\begin{document}
%
\title{Early Acceptance Matching Game for \\ User Association in 5G Cellular HetNets}
%
%
%
\author{\IEEEauthorblockN{Alireza Alizadeh and Mai Vu\\}
\IEEEauthorblockA{Department of Electrical and Computer Engineering,
Tufts University, Medford, USA\\
Email: \{alireza.alizadeh, mai.vu\}@tufts.edu}
\vspace*{-2em}
}


%
%

\markboth{}%
{Shell \MakeLowercase{\textit{et al.}}:  Early Acceptance Matching Game for User Association in 5G Cellular HetNets}
%



\maketitle

\begin{abstract}
In this paper we examine the use of matching theory for user association in millimeter wave (mmWave)-enabled cellular heterogeneous networks (HetNets). 
In a mmWave system, the channel variations can be fast and unpredictable, rendering centralized user association potentially inefficient. 
We propose an efficient distributed matching algorithm, called early acceptance (EA), tailored for user association in 5G HetNets. The effectiveness of the proposed algorithm is assessed by comparing its performance with the well-known deferred acceptance (DA) matching algorithm, in which user association is delayed until the algorithm finishes. 
Numerical results show that the proposed EA matching algorithm reaches a near-optimal solution when compared with a centralized user association, and leads to a more power-efficient and much faster user association process when compared to the distributed DA algorithm.
\end{abstract}


%
\IEEEpeerreviewmaketitle

\section{Introduction}
\thispagestyle{empty}
%
%
%
%
Future cellular heterogeneous networks (HetNets) will be dense because of the coexistence of many base stations (BSs) with different sizes, transmit power and capabilities. As a result, each BS can have a different \textit{quota} (number of users it can serve simultaneously). 
In such a dense network, a challenging problem is user association. This problem intrinsically is  an optimization problem to find the best connection between BSs and user equipment (UEs) to achieve an optimal network performance, and the optimization variables are usually BS-UE associations indicated by zeros and ones (unique association). Although, future UEs may be able to connect to multiple BSs simultaneously (fractional association), such a scheme has a high complexity and increases the overhead. Thus, we focus on unique user association in which each UE can only be associated with one BS at a time. 

In future HetNets, UEs will be able to work in dual connectivity (DC) mode as they can be equipped with a multi-mode modem (e.g. Qualcomm 5G modem) supporting both sub-6 GHz and millimeter wave (mmWave) bands so that association to either a macro cell BS (MCBS) or a small cell BS (SCBS) is possible. This feature enables the UEs to establish their control plane (via MCBS) and data plane (via MCBS or SCBS) separately. Since a LoS between UE and BS is necessary for having a strong mmWave connection, the UE can actively switch its data plane over a SCBS whenever a LoS link with the SCBS is available.


The unique user association results in a complex mixed interger non-linear optimization problem which is usually NP-hard \cite{Andrews}, \cite{Caire}. Heuristic algorithms have been designed to solve this problem and achieve a near-optimal solution \cite{TWC,zalghout2015greedy}. 
These algorithms often require centralized implementation, and as the network size grows, they can suffer from high computational and time complexity.
Matching theory, on the other hand, has been successfully used for solving the problem of user association in cellular networks \cite{semiari2014matching,saad2014college}. 
This low-complexity mathematical framework can be implemented as a distributed game with two sets of players in which each player builds it own preference list and ranks the players of the other set based on the preference list. As such, matching theory is attractive for designing distributed user association. 


\textcolor{black}{
The directional beamforming in 5G mmWave networks makes the user association and network interference highly dependent. In fact, the steered beams along with the associations define the interference in the network. As a result, the network interference depends on user assocaition. In \cite{TWC}, we proposed an interference-aware user association scheme to consider this dependency and designed a centralized algorithm to solve the user association optimization problem. In this paper, we propose a distributed matching algorithm to solve the proposed interference-aware user association problem in 5G cellular HetNets. 
}

\section{System Model}
\textcolor{black}{
We study the problem of user association in a multi-cell HetNet with MCBSs operating at a microwave (sub-6 GHz) frequency band and SCBSs working at a mmWave band. In this section, we introduce our network model as well as channel and signal models.}
\vspace*{-1em}
\subsection{Network Model}
We consider the downlink of a two-tier mmWave-enabled cellular HetNet with $J_M$ MCBS, $J_P$ SCBSs and $K$ UEs. Let $\mathcal{J}=\mathcal{J}_M\cup\mathcal{J}_P= \{1, ..., J\}$ denotes the set of all BSs with $J=J_M+J_P$, and $\mathcal{K}=\{1, ..., K\}$ represents the set of UEs. 
$M_j$ is the number of antennas at BS $j$ and UE $k$ is equipped with a single-antenna module working at sub-6 GHz band, and a UPA antenna of size $N_k$ operating at mmWave band. Each UE $k$ aims to receive $n_k$ data streams from its serving BS such that $1\leq n_k\leq N_k$, where the upper inequality comes from the fact that the number of data streams for each UE cannot exceed the number of its  antennas.
\vspace*{-1em}
\subsection{Microwave and mmWave Channel Models}
In this subsection, we introduce the sub-6 GHz and mmWave channel models.
In the sub-6 GHz band the transmissions are omnidirectional and we use the well-known Gaussian MIMO channel model \cite{Telatar}. 
Thus, we denote $H_{\mu\text{W}}$ as the channel matrix between a MCBS and a UE where the channel entries are i.i.d. complex Gaussian random variables given by $h_{\mu\text{W}} \sim \mathcal{CN}(0,1)$.
In the mmWave band, the transmissions are highly directional and we can not use the simple Gaussian MIMO channel. Instead, we employ the specific clustered mmWave MIMO channel model which includes $C$ clusters with $L$ 
$L$ rays per cluster defined as  \cite{SS}, \cite{Nokia}
\begin{align}\label{clustered_ch}
H_{\text{mmW}}=\frac{1}{\sqrt{CL}}\sum_{c=1}^{C}\sum_{l=1}^{L} \sqrt{\gamma_c}~\mathbf{a}(\phi_{c,l}^{\textrm{UE}},\theta_{c,l}^{\textrm{UE}}) ~\mathbf{a}^*(\phi_{c,l}^{\textrm{BS}},\theta_{c,l}^{\textrm{BS}})
\end{align}
where $\gamma_c$ is the power gain of the $c$th cluster. The parameters $\phi^{\textrm{UE}}$, $\theta^\textrm{UE}$, $\phi^\textrm{BS}$, $\theta^\textrm{BS}$ represent azimuth angle of arrival (AoA), elevation angle of arrival (EoA), azimuth angle of departure (AoD), and elevation angle of departure (EoD), respectively. The vector $\mathbf{a}(\phi,\theta)$ is the response vector of a uniform planar array (UPA) which allows 3D beamforming in both the azimuth and elevation directions \cite{TWC}.
We consider the probability of LoS and NLoS as given in  \cite{RapLetter}, and we utilize the path loss model for LoS and NLoS links as given in \cite{Nokia}.
\vspace*{-1em}
\subsection{Signal Model}
We assume each BS's load is determined based on its available data streams. Thus, we define the total number of downlink data streams sent by BS $j$ as
\begin{equation}\label{D_j}
D_j=\sum_{k \in \mathcal{Q}_j(t)}n_k
\end{equation}
where $\mathcal{Q}_j(t)$ is called the \textit{Activation Set} of BS $j$ which represents the set of active UEs in BS $j$ within time slot $t$, such that  $\mathcal{Q}_j(t) \subseteq \mathcal{K}$, $|\mathcal{Q}_j(t)|=Q_j(t)\leq K$, and $Q_j(t)$ is the quota of BS $j$. Note that the total number of downlink data streams sent by each BS should be less than or equal to its number of antennas, i.e., $D_j \leq M_j$. For notational simplicity, we drop the time index $t$ in definition of $D_j$, and only keep the time index for $Q_j(t)$ due to its importance.

Considering the set of active UEs at BS $j$, the $M_j\times 1$ transmitted signal from the BS can be defined as
\begin{equation}\label{x_j}
\mathbf{x}_j = \mathbf{F}_j \mathbf{d}_j = \sum_{k\in \mathcal{Q}_j(t)}\mathbf{F}_{k,j}\mathbf{s}_k
\end{equation}
where $\mathbf{s}_k\in \mathbb{C}^{n_k}$ is the data stream vector for UE $k$ consists of mutually uncorrelated zero-mean symbols, with $\mathbb{E}\lbrack \mathbf{s}_k\mathbf{s}_k^*\rbrack = \mathbf{I}_{n_k}
$. The column vector $\mathbf{d}_j\in \mathbb{C}^{D_j}$ represents the vector of data symbols of BS $j$, which is the vertical concatenation of the data stream vectors $\mathbf{s}_k,~k\in\mathcal{Q}_j(t)$, such that $\mathbb{E}\lbrack \mathbf{d}_j\mathbf{d}_j^*\rbrack = \mathbf{I}_{D_j}$.
\textcolor{black}{
Matrix $\mathbf{F}_{k,j}\in\mathbb{C}^{M_j\times n_k}$ is the linear precoder (transmit beamforming matrix) for each UE $k$ associated with BS $j$ which separates user data streams,} and $\mathbf{F}_j\in\mathbb{C}^{M_j \times D_j}$ is the complete linear precoder matrix of BS $j$ which is the horizontal concatenation of users' linear precoders.
$\mathbb{E}[\mathbf{x}_j^* \mathbf{x}_j]\leq P_j $ describes the power constraint at BS $j$, where $P_j$ is the transmit power of BS $j$.

\textcolor{black}{The post-processed signal of UE $k$ after performing receive beamforming is given by}
\begin{equation}\label{y_tilde_k}
\tilde{\mathbf{y}}_k = \sum_{j\in \mathcal{J}}\mathbf{W}_k^* \mathbf{H}_{k,j}\mathbf{x}_j + \mathbf{W}_k^*\mathbf{z}_k
\end{equation}
\textcolor{black}{
where $\mathbf{W}_k\in\mathbb{C}^{N_k \times n_k}$ is the linear combiner (receive beamforming) matrix of UE $k$, $\mathbf{H}_{k,j}\in\mathbb{C}^{N_k\times M_j}$ represents the channel matrix between BS $j$ and UE $k$, and $\mathbf{z}_k\in\mathbb{C}^{N_k}$ is the white Gaussian noise vector at UE $k$, with $\mathbf{z}_k\sim\mathcal{CN}(\mathbf{0},N_0 \mathbf{I}_{N_k})$ and $N_0$ is the power spectral density of the noise. 
The presented signal model is applicable for all types of transmit beamforming and receive combining.
}

In MIMO mmWave systems, hybrid (analog and digital) beamforming can be implemented to reduces cost and power consumption of large antenna arrays \cite{SS}. 
In this paper, we employ the SVD beamforming technique to obtain the beamforming matrices at the transmitters and receivers \cite{TWC}. The developed approach, algorithms and insights for user association, however, can be applied to other types of beamforming.


\section{Load Balancing Unique User Association}\label{LBUA_prob}
We follow the mmWave-specific user association model proposed in \cite{TWC} which takes into account the dependency between user association and interference structure in the network. This model is suitable for mmWave systems where the channels are probabilistic and fast time-varying, and the interference depends on the highly directional connections between UEs and BSs.
In this model, during each time slot $t$ the instantaneous CSI remains unchanged such that we can implement per-time-slot unique association. 
Note that the network operator can also choose to implement the proposed user association algorithms per multiple time slots, based on the channel CSI in the first time slot or the averaged channel CSI. Such a choice will lead to a trade-off between user association overhead and resulting network performance. All analysis and results in this paper are for per-time-slot association.

\vspace*{-1em}
\subsection{Problem formulation}
We follow the problem formulation given in \cite{TWC}. In this formulation the \textit{activation vector} $\mathbold{\beta}(t)$ is defined as
\begin{equation}\label{Beta_eq}
\mathbold{\beta}(t)=[\beta_1(t), ..., \beta_K(t)]^T
\end{equation}
where $\beta_k(t)$ is the called the \textit{activation factor} of UE $k$ and represents the index of BS to whom user $k$ is associated with during time slot $t$, i.e., $\beta_k(t)\in\mathcal{J}$ with $k\in\mathcal{K}$ and $t\in\mathcal{T}=\{1, ..., T\}$.  
Considering these definitions, the relationship between the activation set of BS $j$ and the activation factors can be described as
\begin{equation}\label{Q_j}
\mathcal{Q}_j(t) = \{ k: \beta_k(t)=j\}.
\end{equation}

The constraints on the activation factors are
\begin{align}
\sum_{j\in\mathcal{J}} 1_{\beta_k(t)}(j) &\leq 1, ~~\forall k\in \mathcal{K}\label{TFA_cons_1}\\
\sum_{k\in\mathcal{K}}1_{\beta_k(t)}(j). n_k &\leq D_j, ~~\forall j\in \mathcal{J}\label{TFA_cons_2}
\end{align}
where $1_{\beta_k(t)}(j)=1$ if $\beta_k(t)=j$, and $1_{\beta_k(t)}(j)=0$ if $\beta_k(t)\neq j$.
The constraints in (\ref{TFA_cons_1}) reflect the fact that each UE cannot be associated with more than one BS in each time slot (unique association), and the resource allocation constraints in (\ref{TFA_cons_2}) denote that the sum of data streams of UEs served by each BS cannot exceed the total number of available data streams on that BS. We assume that user association is performed during each time slot $t$, and thus we drop the time index $t$ for notational simplicity.




\vspace*{-1em}
\subsection{Instantaneous user rate}
When UE $k$ is connected to BS $j$ in time slot $t$, its instantaneous rate can be obtained as \cite{Telatar,TWC},
\begin{equation}\label{R_kj}
R_{k,j}(\mathbold{\beta}) = \log_2\Big |\mathbf{I}_{n_k} + \mathbf{V}_{k,j}^{-1}(\mathbold{\beta})\mathbf{W}_{k}^*\mathbf{H}_{k,j}\mathbf{F}_{k,j} \mathbf{F}_{k,j}^* \mathbf{H}_{k,j}^*\mathbf{W}_{k}\Big |
\end{equation}
where $\mathbf{V}_{k,j}$ is the interference and noise covariance matrix given as
\begin{align}\label{Y_interf}
&\mathbf{V}_{k,j}(\mathbold{\beta})= \mathbf{W}_{k}^*\mathbf{H}_{k,j}\Big( \sum_{\substack{l\in \mathcal{Q}_{j} \\ l\neq k}} \mathbf{F}_{l,j} \mathbf{F}_{l,j}^* \Big ) \mathbf{H}_{k,j}^*\mathbf{W}_{k}  \nonumber \\
&+ \mathbf{W}_{k}^* \Big( \sum_{\substack{i\in \mathcal{J} \\ i\neq j}} \sum_{\substack{l\in \mathcal{Q}_i}} \mathbf{H}_{k,i}\mathbf{F}_{l,i} \mathbf{F}_{l,i}^* \mathbf{H}_{k,i}^* \Big ) \mathbf{W}_{k} + N_0 \mathbf{W}_k^*\mathbf{W}_k.
\end{align}
Note that the instantaneous rate given in (\ref{R_kj}) is a function of activation sets $\mathcal{Q}_j$. Now, we can express the network \textit{sum-rate} as
\begin{align}\label{r_t}
r(\mathbold{\beta})=\sum_{k\in\mathcal{K}} \sum_{j\in\mathcal{J}} 1_{\mathbold{\beta}_k}(j) R_{k,j}(\mathbold{\beta})
\end{align}
\vspace*{-1.5em}
\subsection{Optimization Problem}
Following \cite{TWC}, the user association optimization problem at each time slot $t$ can be written as
\begin{align}\label{opt_prob2}
\maxi_{\mathbold{\beta}}~~&r(\beta)
\end{align}
subject to the constraints given in (\ref{TFA_cons_1}) and (\ref{TFA_cons_2}). The objective of this optimization problem is to find the optimal activation vector $\mathbold{\beta}$ which maximize the network sum-rate. Here we assume the power of each BS is shared equally among its active UEs, thus the constraint on the transmit power of BSs is automatically satisfied and can be ignored.
\textcolor{black}{
Note that the set of constraints in (\ref{TFA_cons_2}) allows our user association scheme to limit each BS’s load separately. This makes our user association scheme applicable to HetNets where there are different type of BSs with different quotas.
}

The optimization problem in (\ref{opt_prob2}) is a mixed integer non-linear programming (MINLP), which is known to be NP-hard due to its non-convex and nonlinear structure and presence of integer variables. 
In \cite{TWC}, we proposed a centralized efficient algorithm, called \textit{worst connection swapping} (WCS), to find a near-optimal solution for our MINLP. Although, WCS algorithm reaches a near-optimal solution, its complexity increases quadratically as the network size grows. In this paper, we formulate the user association problem as a distributed matching game between BSs and UEs, and design a new matching algorithm to solve our optimization problem. 
\vspace*{-1em}
\section{Matching Theory for Distributed User Association}
In this section, we first model our user association optimization problem as a matching game. Next, we discuss the deferred acceptance game (DA) introduced in \cite{gale1962college}. Finally, we propose a new matching game suitably designed for user association in 5G HetNets. 
\vspace*{-1em}
\subsection{Definitions of User Association Matching Game}
User association problem can be posed as a college admission game where BSs with their specific quota represent colleges and UEs can be considered as students. This framework is suitable for user association in a HetNet where there are different type of BSs with different quotas and capabilities. In order to formulate our user association as a matching game, we first provide some definitions based on the theory of two-sided matching \cite{roth1992two}.

\textbf{Definition 1}: Based on the instantaneous user rates, each UE $k$ (BS $j$) builds a \textit{preference relation} $\succeq_k$ ($\succeq_j$) between each pair of BSs (UEs).

Thus, for any two BSs $i,j \in \mathcal{J},~i\neq j$, we can write
\begin{equation}\label{prf_rlt_K}
j \succeq_k i \Leftrightarrow \Psi_{kj}^{\text{UE}} \geq \Psi_{ki}^{\text{UE}} \Leftrightarrow \text{UE}~k~\text{prefers BS}~j~\text{to BS}~i
\end{equation}
where $\Psi^{\text{UE}}$ is the objective function of UEs, and $\Psi^{\text{UE}}_{kj}$ is the value of objective function when UE $k$ is associated with BS $j$.
Similarly, for any two UEs $k,l \in \mathcal{K},~k\neq l$, each BS builds a preference relation $\succeq_j$ such that 
\begin{equation}\label{prf_rlt_J}
k \succeq_j l \Leftrightarrow \Psi_{kj}^{\text{BS}} \geq \Psi_{lj}^{\text{BS}} \Leftrightarrow \text{BS}~j~\text{prefers UE}~k~\text{to UE}~l
\end{equation}
where $\Psi^{\text{BS}}$ is the objective function of BSs. 

Based on the preference relations, each UE (BS) builds its own \textit{preference list} over the set of all BSs (UEs) in descending order of interest. Note that the length of these preference lists for UEs and BSs are $J$ and $K$, respectively. 
Thus, we can use the preference lists to build a \textit{preference matrix} $\mathbold{P}_{\mathcal{K}}$ of size $K\times J$ for UEs and a \textit{preference matrix} $\mathbold{P}_{\mathcal{J}}$ of size $J\times K$ for BSs. 
Given these preference matrices, we can now define a user association matching game.

\begin{algorithm}[t]
\SetAlgoLined
\KwData{$\mathcal{J}$, $\mathcal{K}$, $\mathbold{P}_{\mathcal{J}}$, $\mathbold{P}_{\mathcal{K}}$, $\mathbold{q}_{\mathcal{J}}$}
\KwResult{Activation vector $\mathbold{\beta}$ }
\textbf{Initialization}: 
Set the preference index $m=1$, and form the rejection vector, $\mathbf{v}_{\text{rej}}=\{1, 2, ..., K\}$\;
\While{$\mathbf{v}_{\textrm{rej}}\neq\varnothing$}{
Each UE $k$ from $\mathbf{v}_{\text{rej}}$ applies to the next (first, if m=1) available BS $j$ (with $Q_j\neq 0$) in $\mathbold{P}_{\mathcal{K}}(k,:)$\;
\If{$k \in \mathbold{P}_{\mathcal{J}}(j,1:Q_j)$}{
$\beta_k=j$\;
$Q_j\leftarrow Q_j-1$\;
\If{$Q_j=0$}{
Remove BS  $j$ from $\mathbold{P}_{\mathcal{K}}$\;
}
Remove UE $k$ from $\mathbf{v}_{\text{rej}}$ and $\mathbold{P}_{\mathcal{J}}$\;
}
$m\leftarrow m+1$\;
}
\caption{Early Acceptance Matching Game}
\end{algorithm}

\textbf{Definition 2}: A user association \textit{matching game} $\mathcal{G}$ is defined by the tuple ($\mathcal{J}$, $\mathcal{K}$, $\mathbold{P}_{\mathcal{J}}$, $\mathbold{P}_{\mathcal{K}}$, $\mathbold{q}_{\mathcal{J}}$) with $\mathbold{q}_{\mathcal{J}}=[Q_1, Q_2, ..., Q_J]$ being the vector of BSs' quotas. 

The outcome of this game is a matching between the set of UEs and the set of BSs, such that each UE $k$ is associated with only one BS (unique association), and each BS $j$ is matched to at most its quota ($Q_j$) of UEs. This matching is defined as follows:

\textbf{Definition 3}: A \textit{matching} $\mu$ is a function which maps an element from $\mathcal{K}\cup \mathcal{J}$ into a subset of elements of $\mathcal{K}\cup \mathcal{J}$ with the following properties
\begin{enumerate}
\item $\mu(k)\subset J$ with $|\mu(k)|=1$ for each UE $k$.
\item $\mu(j)\subseteq K$ with $|\mu(j)|=Q_j$ for each BS $j$.
\item $\mu(k)=j$ if and only if $k\in \mu(j)$.
\end{enumerate}
The last property states the matching $\mu$ is bilateral in the sense that a UE is associated with a BS if and only if the BS accepts that UE. Considering the definition of activation vector $\beta$ in (\ref{Beta_eq}), it can be inferred that the matching $\mu$ on the set of UEs is equal to the activation vector, i.e., $\mu(k) = \beta_k,~\forall k\in \mathcal{K}$.

In a user association matching game, the network nodes (BSs and UEs) are considered as players. Each player may have an objective function (as in (\ref{prf_rlt_K}) and (\ref{prf_rlt_J})) from which it can build its own preference list. 
In the previous works, researchers usually formulate different objective functions for UEs and BSs \cite{semiari2014matching,saad2014college}. This approach needs more information and increases the complexity of computations in practical scenarios. In our proposed matching algorithm, we consider the user instantaneous rate in (\ref{R_kj}) as the objective function for both sides of the game, i.e., $\Psi_{kj}^{\text{UE}}=\Psi_{kj}^{\text{BS}}=R_{kj}(\mathbold{\beta})$. 
This objective function only depends on SINR which can be computed at the UEs and reported to network through physical uplink shared channel (PUSCH) or physical uplink control channel (PUCCH). This makes our proposed matching algorithm very fast and suitable for 5G HetNets which include ultra reliable low latency communications (URLLC).




\begin{algorithm}[t]
\SetAlgoLined
\KwData{$\mathcal{J}$, $\mathcal{K}$, $\mathbold{q}_{\mathcal{J}}$, LoS, CSI, Distances}
\KwResult{Near-optimal activation vector $\mathbold{\beta}$ }
\textbf{Initialization}: Randomly generate initial $\mathbold{\beta}^1$ ($n=1$) according to BSs' quotas\;
\While{$r(\mathbold{\beta^n}) < r(\mathbold{\beta^{n+1}})$}{
Calculate $R_{k,j}(\mathbold{\beta}^n),~\forall k, j$\;
Build preference matrices $\mathbold{P}_{\mathcal{J}}$ and $\mathbold{P}_{\mathcal{K}}$\;
Perform the matching game (DA or EA in Alg. 1) to obtain $\mathbold{\beta}^{n+1}$\;
$n\leftarrow n+1$\;
}
\caption{User Association Matching Alg.} 
\end{algorithm}

\vspace*{-1em}
\subsection{Matching Games}
\subsubsection{Deferred Acceptance Matching Game}
The well-known and widely-used \textit{deferred acceptance} (DA) game is introduced by Gale and Shapley \cite{gale1962college}. 
At the beginning of this game, all UEs apply to their most preferred BS. Each BS $j$ ranks its applicants based on its preference list, keeps the first $Q_j$ UEs in a waiting list, and rejects the rest. Rejected UEs apply to their next preferred BS, and again each BS forms a new waiting list by selecting the top $Q_j$ UEs among the new applicants and those on its previous waiting list. The game terminates when every user has been wait-listed or has applied to all BSs. Note that we assume the BSs have enough quota to accommodate all the UEs in the network. Thus, the rejection list will be empty at the end of deferred acceptance game. 



\subsubsection{Proposed Early Acceptance Matching Game}
When using as a user association game, as it comes from its name, the DA procedure \textit{defers} the association of UEs and BSs to the last iteration of the algorithm due to the fact that all UEs are kept in waiting lists until the last iteration. This game can result in a long delay for the association process and  can be problematic when it comes to user association in a fast varying mmWave system. 

In order to overcome the aforementioned problem, we propose a new matching game, called \textit{early acceptance} (EA) game. 
In the EA game, BSs immediately decide about the acceptance or rejection of applicants which result in a faster and more efficient user association procedure. The EA game starts with forming a rejection vector which includes all UEs. In the first iteration of the game, each UE $k$ applies to BS $j$ if that BS is the most preferred one for the UE, i.e.,
\begin{equation}
j \succeq_k i,~i\in \mathcal{J}.
\end{equation}
Then, each BS $j$ immediately accepts those UEs which are among its first $Q_j$ preferred UEs, i.e., in $j$th row of  $\mathbold{P}_{\mathcal{J}}$. When a UE $k$ is accepted by BS $j$, we update the association vector with this connection ($\beta_k=j$), and then the BS updates its quota ($Q_j\leftarrow Q_j-1$). Since the UE has been accepted, it will be removed from the rejection vector and also from the preference matrix  $\mathbold{P}_{\mathcal{J}}$. Similar to the DA game, if a UE is rejected by a BS, it will be kept in the rejection vector and apply to its next preferred BS in the next iteration of the game. 
In the following iterations, each UE (from the rejection vector) applies to its next preferred BS with $Q_j\neq 0$.
Whenever a BS is ran out of quota, it will be removed from the preference matrix $\mathbold{P}_{\mathcal{K}}$, and thus no UE will apply to that BS again. A summary of the EA game is described in Algorithm 1. 

The solution of the matching game associates each UE $k$ to a BS $j$, and it determines the set of UEs associated with each BS $j$. As discussed earlier, given the matching $\mu$, the activation vector $\beta$ can be obtained easily. Thus, in what follows we use activation vector $\beta$ as the solution of our matching algorithm. 
\begin{figure}[t]
\centering
\vspace*{-1em}
\includegraphics[scale=.33]{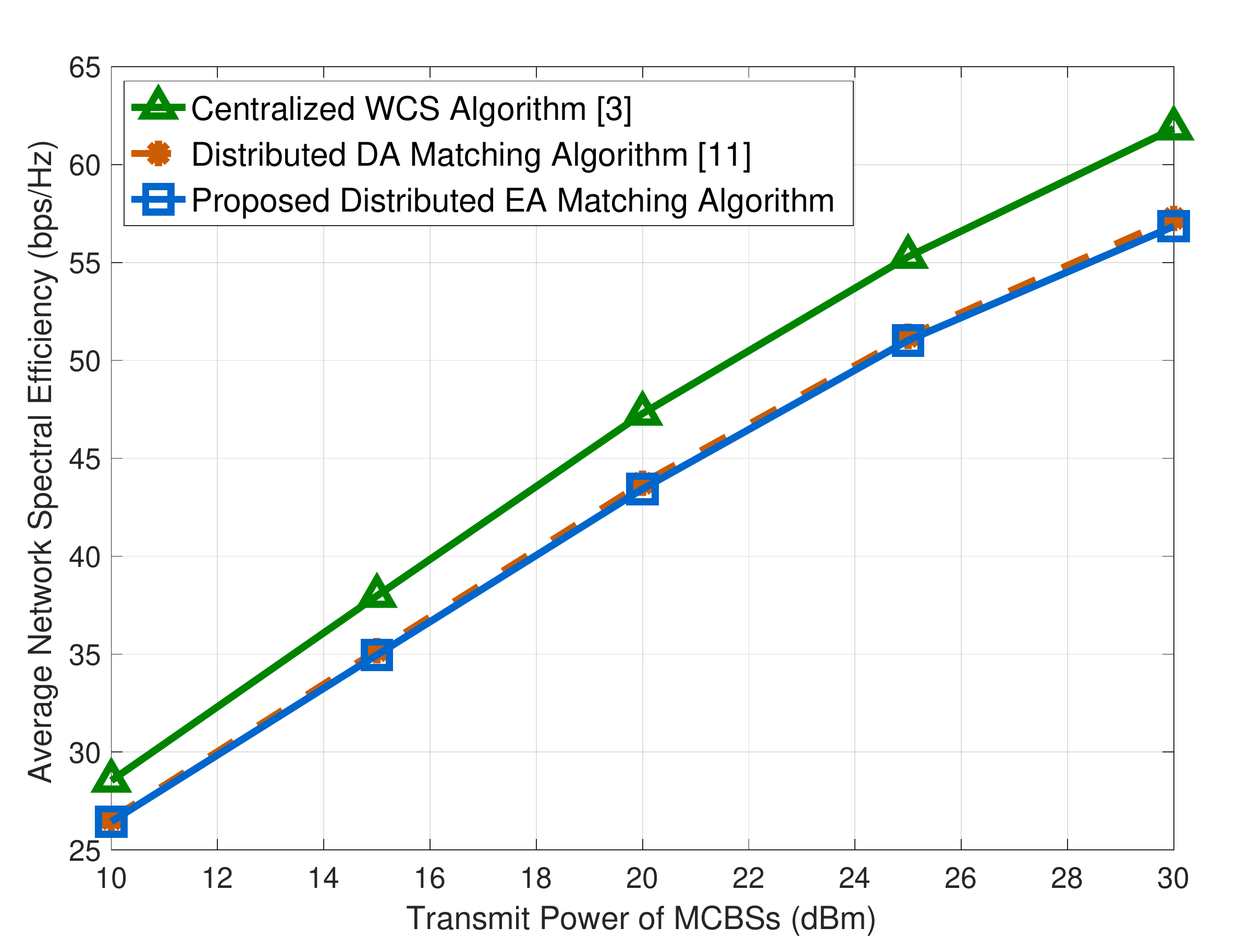}
\vspace*{-.8em}
\caption{Average network spectral efficiency for centralized and distributed load balancing user association schemes}
\label{SumRate}
\end{figure}
\vspace*{-1em}
\subsection{User Association Matching Algorithm}
The matching algorithm starts with a randomly generated association vector $\mathbold{\beta}^1$ based on the BSs' quotas to satisfy the BSs' load constraints in (\ref{TFA_cons_2}). At each iteration, considering the current activation vector $\beta^n$, each UE computes the instantaneous rates it can get from each BS and reports these rates to the BSs via PUSCH or PUCCH as discussed in Section IV.A. Then, the BSs and UEs update their preference lists based on the rate information. Next, we need to perform a matching game (DA or EA) to obtain the new activation vector $\beta^{n+1}$. The algorithm stops when the network sum-rate does not improve, i.e., $r(\mathbold{\beta^n}) \geq r(\mathbold{\beta^{n+1}})$.
The process is summarized in Algorithm 2.

In the next section, we compare the performance of our proposed distributed EA algorithm with the DA algorithm and also with the centralized WCS algorithm introduced in \cite{TWC}.
We compare the objective of the network sum-rate in (\ref{r_t}) achieved by all three algorithms. For the distributed matching algorithms, while matching stability is an important consideration in a game without an explicit objective \cite{gale1962college}, here since we have a specific objective for the game and, furthermore, the matching is per-time-slot and is likely to change each time user association is performed, matching stability consideration can be irrelevant. Instead, we consider other metrics relevant to the distributed user association problem. 
In particular, we compare the two centralized matching algorithms in terms of the following important metrics:
\begin{figure}[t]
\centering
\vspace*{-1em}
\includegraphics[scale=.33]{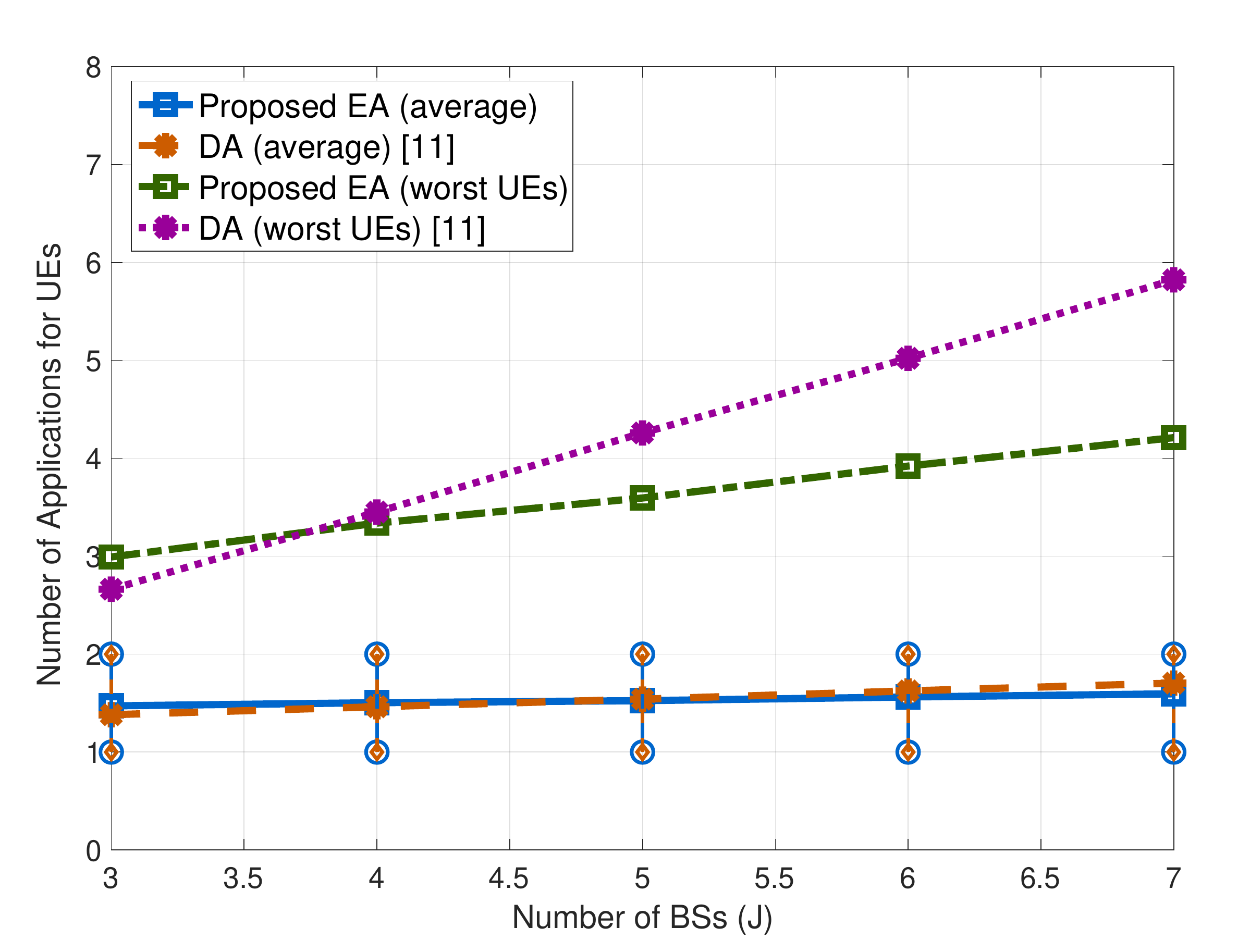}
\vspace*{-.8em}
\caption{Comparing users' worst case and average number of applications with 25 and 75 percentile bars for EA and DA algorithms}
\label{Appl}
\end{figure}

1. \textit{User's number of applications}: During a user association matching game, each UE applies to its preferred BSs, until it is accepted by one of them. Each application requires a signaling interaction between the UE and BS, which is a power-consuming process. Thus, the lower the number of applications, the less the power consumption at the UE. 

2. \textit{Users' acceptance delay}: This metric is represented by number of iterations until a UE is accepted (associated). As stated earlier, in the DA game, the association of all UEs is postponed to the last iteration of the algorithm. Thus, the acceptance delay for all UEs is the same. In contrast, in the EA game, the acceptance delay for each UE is equal to the number of its applications. 

\vspace*{-1em}
\section{Numerical Results}\label{Sim_res}
In this section, we evaluate the performance of the proposed user association matching algorithm in the downlink of a 5G HetNet with $J$ BSs and $K$ UEs. The network includes 1 MCBS operating at 1.8 GHz and $J-1$ SCBSs operating at 73 GHz. 
The channel elements for sub-6 GHz links and the mmWave links are generated as described in Sec. II-B. We assume each mmWave link is composed of 5 clusters with 10 rays per cluster. In order to implement 3D beamforming, each BS is equipped with a UPA of size $8\times 8$, and each UE is equipped with a single-antenna module designed for sub-6 GHz band, and a $2\times 2$ UPA of antennas designed for mmWave band. 
Also, we assume that the transmit power of MCBS is 10 dB higher that the one for SCBSs.
Network nodes are deployed in a $300 \times 300~\textrm{m}^2$ square where the BSs are placed at specific locations and the UEs are distributed randomly according to a uniform distribution.



Fig. \ref{SumRate} compares the spectral efficiency of a HetNet with 1 MCBS, 4 SCBSs, and 24 UEs ($Q_1=8$, $Q_j=4, j=2,...,5$) for three association schemes: 1) WCS algorithm \cite{TWC}, 2) DA matching algorithm, and 3) EA matching algorithm. This figure shows that while the centralized WCS algorithm slightly outperforms the distributed matching algorithms as expected, both distributed algorithms achieve about 92\% the rate achieved by the centralized one. Although we did not include a complexity comparison due to limited space, we observed that the distributed algorithms have lower complexity and they are much faster than the centralized WCS algorithm. 


\begin{figure}[t]
\centering
\vspace*{-1em}
\includegraphics[scale=.33]{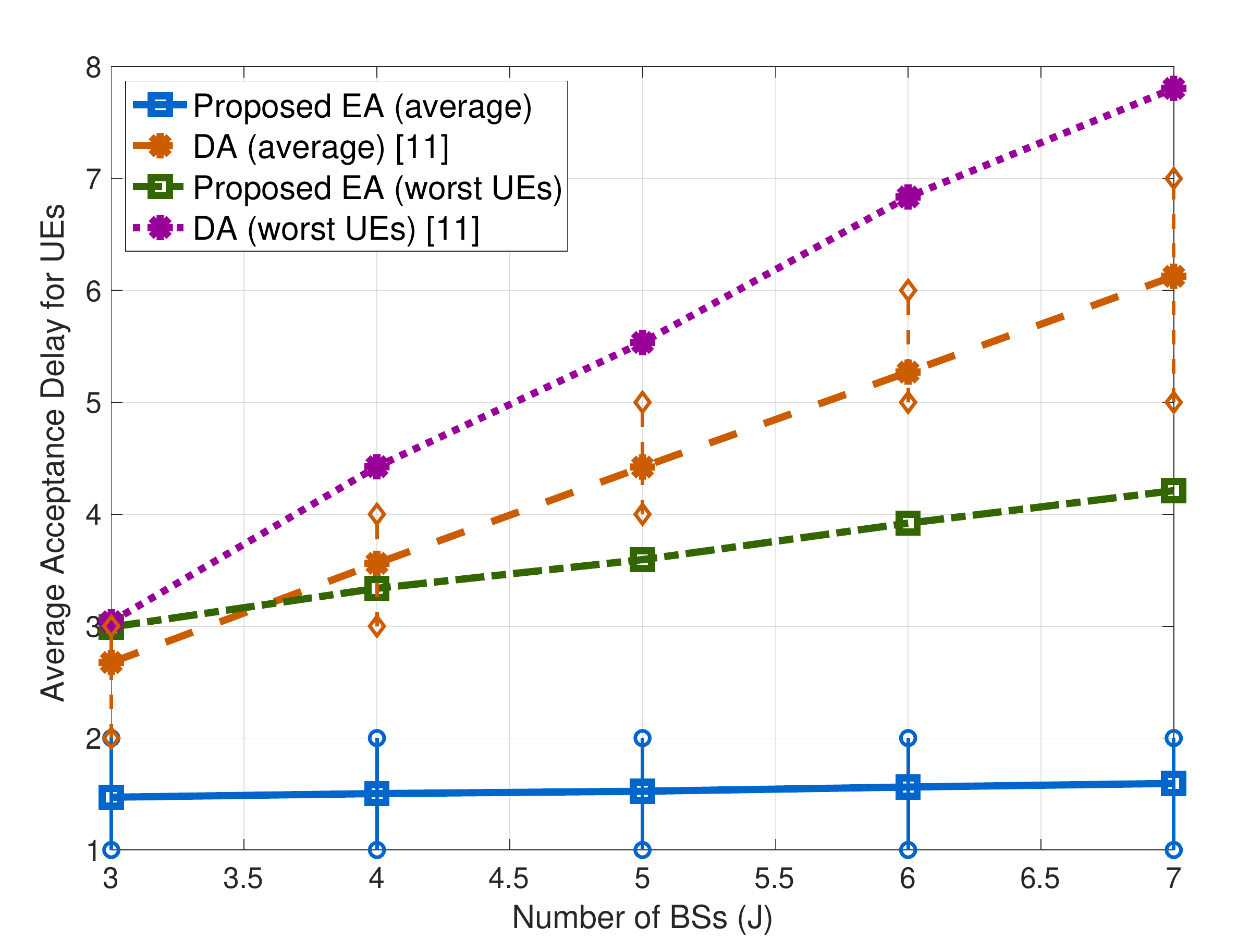}
\vspace*{-1em}
\caption{Comparing users' worst case and average acceptance delay with 25 and 75 percentile bars for EA and DA matching algorithms}
\vspace*{-1em}
\label{Delay}
\end{figure}
For the next simulations, we increase the number of BSs and UEs while keeping the BSs' quota fixed ($Q_1=6$, $Q_j=3, j=2,...,J$). The average is taken over 200 random UEs' locations and 10 channel realizations per UEs' location. 
Fig.  \ref{Appl} compares the EA and DA algorithms in terms of users' average number of applications, and average number of applications for worst users (which have the highest number of applications). This figure shows that the algorithms have a similar performance in terms of user's average number of applications, but in the proposed EA algorithm the worst users have a lower number of applications, which means that the EA algorithm is more power-efficient than the DA algorithm.

Fig. \ref{Delay} shows that the EA algorithm significantly outperforms the DA algorithm in terms of users' acceptance delay in both the average and worst cases. In most cases, the worst case delay for EA is even better than the average delay for DA. 
Thus, we can conclude the EA game results in a faster association process. This advantage becomes more significant as the network size grows. 



\begin{figure}[t]
\centering
\vspace*{-1em}
\includegraphics[scale=.33]{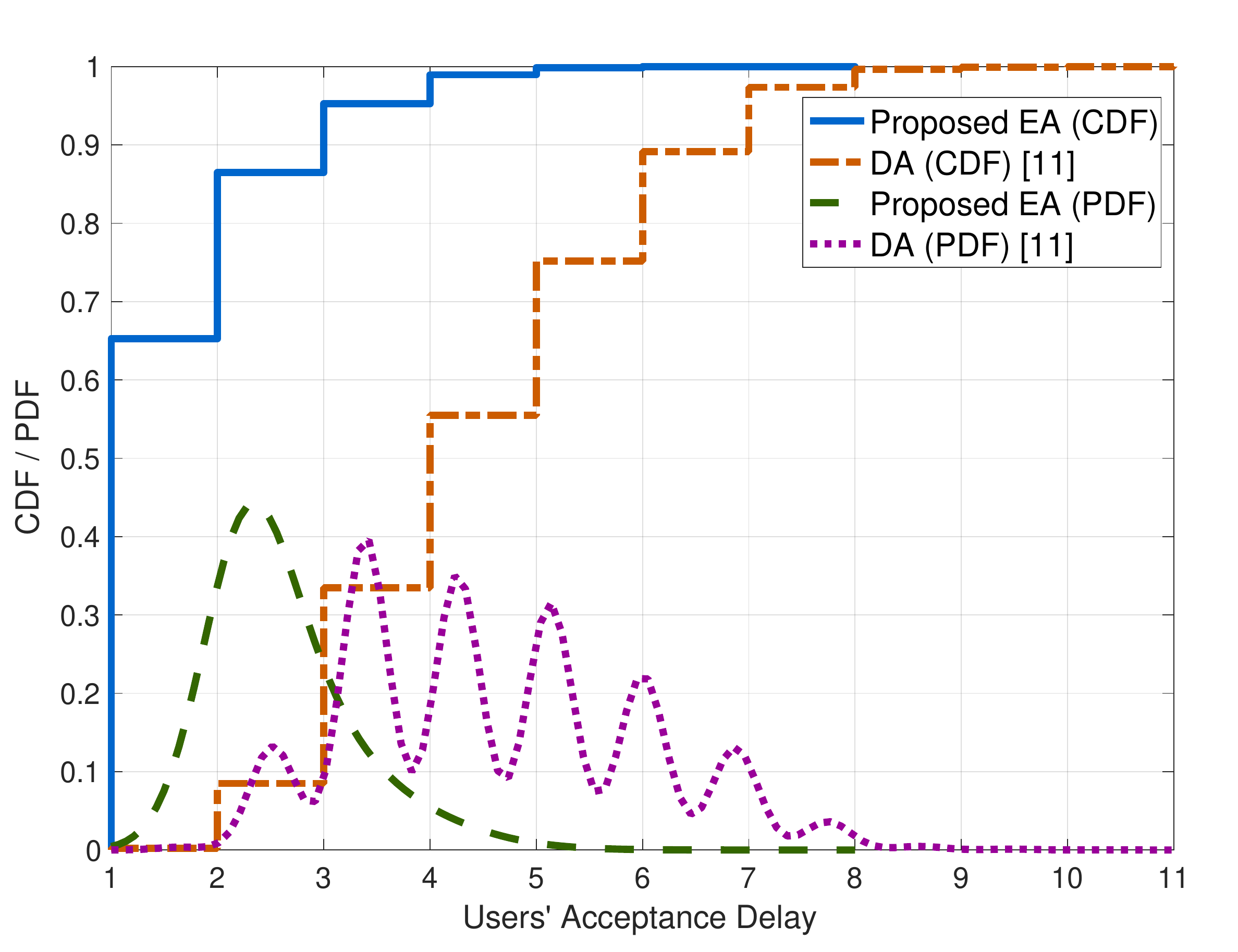}
\vspace*{-1em}
\caption{Comparing the CDF and PDF of users' acceptance delay for EA and DA matching algorithms}
\vspace*{-1em}
\label{CDF_PDF_Delay}
\end{figure}

Fig. \ref{CDF_PDF_Delay} depicts the CDF and PDF of users' acceptance delay for the user association matching algorithms. It can be seen from the figure that the probability of having less acceptance delay is much higher for the EA algorithm, confirming the fact that the association process in the proposed EA algorithm is much faster compared to the DA algorithm while achieving the same network throughput.

\vspace*{-1em}
\section{Conclusion}
We proposed a distributed EA matching algorithm for user association in 5G HetNets, and compared its performance with the well-known DA matching algorithm and the centralized WCS algorithm. We showed that the proposed EA algorithm achieves a network throughput close to that of the centralized WCS algorithm, while incurring a much lower complexity and overheads due to its distributed nature. Considering users' number of applications which affects the power consumption and users' acceptance delay metrics, we showed that the proposed EA algorithm is more power-efficient and results in a faster association process compared to the well-known DA algorithm while achieving the same network throughput. These results suggest that EA may be more suitable for real-time user association in a wireless network. 

%

\ifCLASSOPTIONcaptionsoff
  \newpage
\fi



%

\bibliographystyle{IEEEtran}
\bibliography{References_etal}

%









\end{document}